# The Power Ratio and the Interval Map:
## Spiking Models and Extracellular Recordings

Daniel S. Reich[1,2], Jonathan D. Victor[1,2], and Bruce W. Knight[1]

[1]Laboratory of Biophysics, The Rockefeller University, 1230 York Avenue, New York, NY 10021

[2]Department of Neurology and Neuroscience, Cornell University Medical College, 1300 York Avenue, New York, NY 10021

<u>Address correspondence to:</u>
Daniel Reich
The Rockefeller University
1230 York Avenue, Box 200
New York, NY 10021
Tel: 212-746-6520
Fax: 212-746-8389
E-mail: reichd@rockvax.rockefeller.edu

<u>Abbreviated Title</u>
Power ratios and interval maps

<u>Key Words</u>
spike trains, retinal ganglion, lateral geniculate nucleus, primary visual cortex, neural models, neural noise, temporal coding, rate coding, Poisson process, renewal process, refractory period, integrate-and-fire, interval distributions

<u>Acknowledgments</u>
The authors wish to thank Mary Conte, Rob de Ruyter van Steveninck, Ehud Kaplan, Ferenc Mechler, Pratik Mukherjee, Tsuyoshi Ozaki, Keith Purpura, Mavi Sanchéz-Vives, Niko Schiff, and Haim Sompolinsky. Supported by NIH grants GM07739 and EY07138 (DSR) and EY9314 (JDV).

**Abstract**

We describe a new, computationally simple method for analyzing the dynamics of neuronal spike trains driven by external stimuli. The goal of our method is to test the predictions of simple spike-generating models against extracellularly recorded neuronal responses. Through a new statistic called the *power ratio*, we distinguish between two broad classes of responses: (1) responses that can be completely characterized by a variable firing rate, (for example, modulated Poisson and gamma spike trains); and (2) responses for which firing rate variations alone are not sufficient to characterize response dynamics (for example, leaky integrate-and-fire spike trains as well as Poisson spike trains with long absolute refractory periods). We show that the responses of many visual neurons in the cat retinal ganglion, cat lateral geniculate nucleus, and macaque primary visual cortex fall into the second class, which implies that the pattern of spike times can carry significant information about visual stimuli. Our results also suggest that spike trains of X-type retinal ganglion cells, in particular, are very similar to spike trains generated by a leaky integrate-and-fire model with additive, stimulus-independent noise that could represent background synaptic activity.

**Introduction**

A central issue in neuroscience is the question of whether neuronal spike trains *in vivo* are essentially random (Shadlen and Newsome, 1994; Shadlen and Newsome, 1998) or have temporal structure that might convey information in some form other than the mean firing rate (Rieke et al., 1997). Evidence is accumulating in favor of the second hypothesis (Abeles et al., 1994; Hopfield, 1995; Singer and Gray, 1995): temporal codes have been found in the discharges of individual neurons both *in vitro* (Mainen and Sejnowski, 1995; Nowak et al., 1997) and *in vivo* (Cattaneo et al., 1981; Mandl, 1993; Richmond and Optican, 1987; Victor and Purpura, 1996; Mechler et al., 1998). Moreover, spike timing in the visual cortex of monkeys has a well-structured relationship to elementary features of visual stimuli, such as orientation, contrast, and spatial frequency (Victor and Purpura, 1997).

In some retinal ganglion cells and lateral geniculate nucleus (LGN) relay neurons, spike timing is sufficiently precise to be manifest as discrete peaks in peri-stimulus time histograms (PSTHs) (Berry et al., 1997; Reich et al., 1997; Tzonev et al., 1997). This result is consistent with a model that treats retinal ganglion cells as noisy, leaky integrate-and-fire (NLIF) devices, but it is also consistent with simpler models that treat retinal ganglion cell spike trains as modulated Poisson processes for which the measured PSTH is an estimate of the time-varying probability density for spike firing.

We present a powerful method for distinguishing between two broad classes of models. The first class, which we call "simply modulated renewal processes" (SMRPs), gives responses that can be completely characterized as renewal processes with varying firing rates. The second class, by contrast, has dynamics that induce patterning of spike times and spike intervals in a stimulus-dependent manner. Fundamentally, the two classes of models differ in the way the underlying spike generating mechanism interacts with external stimuli. Here, we show that spike trains of most neurons in the early stages of the mammalian visual system cannot be modeled as SMRPs.

**Materials and Methods**

*Recordings.* We made extracellular recordings of the activity of LGN neurons and their retinal inputs in anesthetized cats. We also recorded the activity of V1 neurons in anesthetized macaque monkeys. Experiments were performed on 9 male and 3 female adult cats, and on 3 male adult monkeys, which weighed roughly 3 kg each. All experimental procedures complied with the National Eye Institute's guidelines, *Preparation and Maintenance of Higher Mammals During Neuroscience Experiments*. Portions of this work were presented at the 1998 FASEB conference, *Retinal Neurobiology and Visual Perception*. In addition, a small fraction of the data, analyzed in different ways, has been published elsewhere (Reich et al., 1997).

For the cats, anesthesia was initiated by intramuscular injections of xylazine 1 mg/kg (Rompun, Miles, Shawnee Mission, KS) and ketamine 10 mg/kg (Ketaset, Fort Dodge, Fort Dodge, IA) and was maintained throughout surgery and recording with intravenous injection of thiopental 2.5%, 2-6 mg$\cdot$kg$^{-1}\cdot$hr$^{-1}$ (Pentothal, Abbott, Abbott Park, IL). Paralysis was induced and maintained with vecuronium 0.25 mg$\cdot$kg$^{-1}\cdot$hr$^{-1}$ (Norcuron, Organon, West Orange, NJ). For the monkeys, anesthesia was induced with ketamine 10 mg/kg, supplemented as needed by methohexital 0.5-1 mg/kg (Brevital, Eli Lilly, Indianapolis) boluses during the preparatory surgery, and maintained with sufentanil 3 µg/kg bolus, 1-6 µg/kg/hr (Sufenta, Janssen, Titusville, NJ). Paralysis was induced and maintained with pancuronium 1 mg bolus, 0.2-0.4 mg/kg/hr (Pavulon, Elkins-Sinn, Cherry Hill, NJ).

Gas-permeable hard contact lenses were used to prevent corneal drying, and artificial pupils (3 mm diameter) were placed in front of the eyes. The optical quality of the animals' eyes was checked regularly by direct ophthalmoscopy. Optical correction with trial lenses was added to optimize grating responses at a viewing distance of 114 cm. Blood pressure, heart rate, expired carbon dioxide, and core temperature were continuously monitored and maintained within the physiological range.

Tungsten-in-glass electrodes (Merrill and Ainsworth, 1972) recorded extracellular potentials from individual cat LGN neurons and from their primary retinal ganglion cell inputs in the form of synaptic (S) potentials (Kaplan et al., 1987), or from monkey V1 neurons. The electrode signals were amplified, filtered, and monitored conventionally. Action potentials of single neurons were selected by a window discriminator (Winston Electronics, Millbrae, CA) for the cats. For the monkeys, analog waveforms were identified and differentiated on the basis of criteria such as peak amplitude, valley amplitude, and principal components (Datawave, Longmont, CO). Visual stimuli were created on a white CRT (Conrac 7351, Monrovia, CA, 135 frames/s, 80 cd/m$^2$ mean luminance) for the cats, or on a green CRT (Tektronix 608, Wilsonville, OR, 270 frames/s, 150 cd/m$^2$ mean luminance) for the monkeys, by specialized equipment developed in our laboratory. Action potentials were timed to the nearest 0.1 ms.

Cat retinal ganglion and LGN neurons were classified as X-type or Y-type and on-center or off-center (Enroth-Cugell and Robson, 1966). We measured spatial frequency tuning and contrast response functions with drifting sinusoidal gratings, and we sampled between six and 10 separate contrasts or spatial frequencies for each neuron. We usually recorded the responses to each contrast or spatial frequency for 16 seconds, but occasionally for longer periods of time (up to 256 s), before the next stimulus was presented.

Monkey V1 neurons were classified as simple or complex on the basis of whether their response to a drifting grating of high spatial frequency was predominantly a modulated response at the driving frequency, for simple cells, or else an elevation of the mean firing rate, for complex cells (Skottun et al., 1991). We measured contrast response functions with sinusoidal gratings presented at the optimal orientation, spatial frequency, and temporal frequency. The stimuli were presented for 4 to 10 seconds at each contrast, in random order, and the entire set of contrasts was presented, in different random orders, four to eight times. For the analysis described in this paper, we considered the responses to each stimulus to be one continuous steady-state record.

*Poisson spike trains*. We used a resampling procedure (Victor and Purpura, 1996) to create artificial spike trains with the same PSTH as a measured spike train. Each spike in the original spike train was associated with a randomly chosen response cycle, an operation that preserved the set of spike times (and hence the PSTH) but destroyed the distribution of those times among the individual cycles (and hence the interspike interval histogram, or ISIH). The resulting spike train had the statistics of a modulated Poisson process.

*Modified Poisson spike trains.* To test the hypothesis that firing rate is in part determined by slow variations in responsiveness, and that such slow variations could account for any difference between recorded data and Poisson-resampled data, we performed a procedure equivalent to the "exchange resampling" of Victor and Purpura (1996). Each response cycle was assigned the same number of spikes as had occurred in the original spike train, but the spike times themselves were drawn at random from the entire collection of spikes. All spikes were used exactly once, and the PSTH of the resampled spike train was therefore identical to the PSTH of the original spike train.

*Gamma spike trains*. We also generated artificial spike trains with similar (though not identical) PSTHs to those of measured spike trains, but with the interval statistics of $n^{th}$-order modulated gamma processes. Gamma processes may be considered to have a relative refractory period, the duration of which changes with the stimulus strength. For very high-order gamma processes, the firing is clock-like and approaches the behavior of a non-leaky integrate-and-fire model. Gamma processes have been suggested as reduced descriptions of retinal ganglion cell spike-generating mechanisms (FitzHugh, 1958; Troy and Robson, 1992). To generate modulated gamma spike trains, we drew a random number to determine whether a spike was fired in each 0.1 ms time bin, where the probability for spike firing in each bin was determined from the linearly-interpolated PSTH. For an $n^{th}$-order gamma process, the model was given $n$ chances to fire in each bin. However, only every $n^{th}$ spike was kept in the final spike train.

*Spike trains with fixed absolute refractory periods*. To generate spike trains with absolute refractory periods, we modified the $n^{th}$ order gamma model so that the firing probability was held at zero for a fixed time, equal to the desired refractory period, after each spike. This procedure effectively shifted the overall ISIH to the right, leaving a gap equal in duration to the refractory period.

*NLIF model.* We used a noisy variation of the leaky integrate-and-fire model (Knight, 1972). This model is a highly reduced version of the Hodgkin-Huxley equations for neuronal firing, in which the state variable $V(t)$ plays the role of the membrane potential. The model "fires" when $V(t)$ reaches a threshold $V_{th}$, after which $V(t)$ is reset to zero. In our simulations, the input to the model was a sinusoidally-modulated current. Poisson-distributed noise shots of steady rate, uniform size, and random polarity were added to the state variable at each time 0.1 ms step. In the absence of noise, this model phase locks: if the leak rate is sufficiently fast compared to the stimulus cycle, and the stimulus sufficiently strongly modulated, the spike times in all stimulus cycles are identical.

Formally, the model is:

$$\frac{dV}{dt} = -\frac{V(t)}{\tau} + \left[ S_0 + S_1 \cos(2\pi f t + \varphi) \right] + N(t)$$

where $V(t)$ is the state variable of the model

| | | |
|---|---|---|
| $t$ | is the time within the stimulus cycle (s) | |
| $\tau$ | is the time constant of the leak (s) | |
| $S_0$ | is the mean input level (s$^{-1}$) | |
| $S_1$ | is the contrast (s$^{-1}$) | |
| $f$ | is the temporal frequency (Hz) | |
| $\varphi$ | is the phase (radians) | |
| $N(t)$ | is the input Poisson shot-noise (s$^{-1}$) | |

The overall firing rate of the neuron depends on the threshold $V_{th}$, the noise, and the deterministic input. We calculated the model's responses to stimuli of 10 different contrasts about a mean of $S_0=1$ s$^{-1}$, ranging from 0% ($S_1=0$ s$^{-1}$) to 100% ($S_1=1$ s$^{-1}$). We used a threshold that was 75% of the steady-state value of the state variable in the absence of input modulation, a time constant $\tau$ of 20 ms, a temporal frequency $f$ of 4.2 Hz, and a phase $\varphi$ of $\pi$ radians, which aligned the period of strongest firing with the middle of the response cycle. We tested several different noise shot sizes ranging from 0 to ±0.0016, but the shot rate was kept constant at 1000 shots/s.

The state variable $V(t)$ was measured in dimensionless units, following Knight (1972). These units can be considered voltages, since the state variable loosely corresponds to the membrane potential of real neurons. However, because we did not use the NLIF model to describe the detailed biophysical processes that occur in real neurons, we chose to retain the original dimensionless units for $V(t)$. Despite the difference in units, our model is similar, in many ways, to the one described by Shadlen and Newsome (1998). The primary difference is that they did not provide their model with a deterministic input, but rather used only the shot-noise process. The deterministic input in our model enabled us to use fewer, smaller-amplitude noise shots. Even so, the noise in our model caused substantial jitter in spike timing, while it dominated the response statistics in the model of Shadlen and Newsome.

## **Results**

After a brief discussion of renewal processes, we describe a multi-step procedure for classifying neuronal responses into one of the two classes mentioned in the Introduction. The first step of this procedure is to apply a data-driven time transformation that flattens the PSTH and converts SMRPs into unmodulated renewal processes. The second step is to plot the distribution of the interspike intervals on the transformed timescale. The final step is to calculate an index that is sensitive to variations in the interspike interval distribution, and to compare that index to the one obtained from Poisson processes with the same PSTH.

*Renewal processes*. Spike trains of renewal processes are characterized by the fact that all interspike intervals are independent and identically distributed (Papoulis, 1991). This implies that the firing rate is necessarily constant, on average. We can write the probability that a spike is fired within a brief time window *dt* at a particular time *t* since the previous spike as

$$p(t)dt = rg(rt)dt$$

where *r* is the firing rate and *g* is some dimensionless function that integrates to 1. This function *g* describes the shape of the interspike interval distribution from which successive spikes are drawn at random. For a Poisson process, the simplest renewal process, *g* is exponential, so

$$p(t)dt = re^{-rt}dt.$$

Real neuronal spike trains have variable firing rates, so we need to relax the strict definition of a renewal process to account for this. We eliminate the requirement that all interspike intervals be identically distributed, but we maintain the requirement that the intervals be independent. Thus, intervals may depend on the stimulus, but they do not reflect the firing history before the previous spike. The effect of our modification is to create a "modulated renewal process" for which the firing probability now depends on a variable firing rate *r(t)* and on an interval distribution that changes in time. Thus, the probability that a spike at time $t_0$ is followed by a spike in time window *dt* at time $t_0+t$ can now be written as $p(t|t_0)dt$.

*Time transformation*. In order to compare responses to different stimuli, we apply a "demodulation" transformation. This time transformation replaces the original time axis by the integral of the PSTH (FitzHugh, 1957; Gestri, 1978; Cattaneo, Maffei, and Morrone, 1981). For each real time *t*, we obtain a transformed time *u(t)* by the following relation:

$$u(t) = \int_0^t \frac{r(t')}{\bar{r}} dt'$$

where *r(t)* is the firing rate at time *t*, estimated by the PSTH, and $\bar{r}$ is the mean firing rate over the entire cycle. This invertible transformation effectively expands time during portions of the response when the firing rate is high and compresses time when the firing rate is low, so that the PSTH in transformed time is flat. The transformation changes the internal clock of the neuron from one that ticks in units of real time into one that ticks in units of instantaneous firing probability. Across all response cycles, the same number of spikes is fired in each unit of transformed time.

Because spike trains are inherently discontinuous, we can implement a computationally simple version of the transformation. To determine the transformed time of a given spike, we multiply the fraction of spikes (across all cycles) that occurred before that spike by the cycle duration, and we break ties randomly. Fig. 1 shows the effects of the time transformation for data derived from the NLIF model at 100% contrast and shot size 0.0004.

*Time transformation and renewal processes*. We now identify a subset of modulated renewal processes, which we call "simply modulated renewal processes" (SMRPs). The spike trains of SMRPs are uniquely converted, by our time transformation, into spike trains of unmodulated renewal processes with the same mean firing rate (Gestri, 1978). Examples of SMRPs are modulated Poisson and gamma processes. On the other hand, examples of modulated renewal processes that are not SMRPs are the NLIF model as well as models with fixed absolute refractory periods (Table 1). These non-SMRP models contain parameters, such as the leak time and the refractory period, that are fixed in real time and not affected by external stimuli or firing rate variations. In transformed time, however, the parameters are no longer fixed, because they are scaled by the local firing rate. As explained below (*Specificity of the power ratio*), this implies that the interspike intervals are not identically distributed in transformed time, as they would be for a true renewal process. Hence, these models are not SMRPs.

*Interval maps*. To distinguish between SMRPs and other models that could have been responsible for a measured spike train, we plot the *interval map*. The interval map relates each spike time (plotted on the horizontal axis) to the subsequent interspike interval (plotted on the vertical axis). Examples of interval maps are shown in Figs. 2A (real time) and 2B (transformed time). The left column uses a spike train generated by the NLIF model, while the right column uses a Poisson-resampled spike train with the same PSTH (see Methods). The interval map is reminiscent of the "intervalogram" of Funke and Worgötter (1997), but there is no binning or averaging. It includes all the information necessary to reconstruct both the PSTH and ISIH of a given dataset. To obtain the PSTH, we simply add up the number of points in each time bin along the horizontal axis, and to obtain the ISIH, we add up the number of points in each time bin along the vertical axis. In all the interval maps in this paper, the PSTH is plotted above the interval map, and the ISIH, rotated 90 degrees, on the righthand side.

In Fig. 2A, the distinct cluster of points at the end of the response cycle in both interval maps corresponds to the final interval in each cycle. In real time, these final intervals, which span the portion of the stimulus cycle during which no spikes were fired (that is, when the PSTH is zero), are far longer than the other intervals. This is true for both NLIF and Poisson spike trains, because the two spike trains have identical PSTHs. In transformed time, however, spikes are equally likely to be fired at all points in the stimulus cycle, so the PSTH is never zero, and there is therefore no long-interval cluster. For a Poisson process in particular, the distribution of interspike intervals is largely independent of transformed time, so the transformed-time interval map of the Poisson-resampled spike train is nearly uniform (Fig. 2B, right panel). In other words, there are no privileged spike times for a Poisson process.

For the NLIF spike train, however, the long-interval cluster is clearly retained (Fig. 2B, left panel, large arrow), indicating that the distribution of interspike intervals is not independent of transformed time, and that some spike times are, in fact, privileged over others. The explanation for this lies in the leakiness of the NLIF model, which ensures that if the external stimulus is sufficiently small (for example, the negative phase of a high-contrast sinusoid), the state variable falls to near its minimum before beginning to recharge as the stimulus grows. This resynchronizes the state variable and causes the first spike in each cycle to occur at a highly reliable, privileged time, regardless of the time of the previous spike.

Thus, the first spike in each cycle is independent of the last spike in the previous cycle for the NLIF spike train, whereas for the Poisson process, the two spike times depend strongly on one another. This is opposite to the relationship between the final interspike interval and the last spike time in each cycle, which are correlated for the NLIF spike train and independent for the Poisson process. In other words, for the NLIF spike train, because of the resynchronization, the final interval is longer when the last spike occurs relatively early, and shorter when the last spike occurs relatively late. The time transformation does not eliminate this dependence, which is reflected in the distinct final-interval cluster (large arrow). To a lesser extent, the leakiness and resetting properties of the NLIF model are also reflected in the smaller interval clusters that occur throughout the cycle (small arrows).

To quantify the extent to which the interval map of a particular spike train deviates from that of a Poisson process with the same PSTH, we use a power-spectral approach to detect the absence (for a Poisson process) or presence (for spike trains that could not have been generated by a Poisson process) of slow changes in the interval map across transformed time. Specifically, we calculate the sample power in the transformed-time interval map at each harmonic of the stimulus cycle, normalized by the total sample power of the modulated (non-DC) harmonics (Fig. 2C). The prominent clusters visible in the transformed-time interval map of the NLIF spike train selectively increase the sample power in the low-frequency harmonics (Fig. 2C, left panel). We therefore focus on the sample power in the first *n* harmonics, where *n* is the mean number of spikes in each response cycle, rounded up to the next integer. These harmonics are signified by circles in Fig. 2C, where the firing rate is bracketed by the pair of solid circles.

We define the *power ratio* as the mean sample power in the first *n* frequency components of the interval map divided by the mean sample power of all modulated (non-DC) components, or

$$PR = \frac{\frac{1}{n}\sum_{k=1}^{n}|H_k|^2}{\frac{2}{N}\sum_{k=1}^{N/2}|H_k|^2}$$

where *N* is the total number of interspike intervals and $H_k$ is the discrete Fourier component at the $k^{th}$ harmonic of the transformed-time interval map. The discrete Fourier components are given as

$$H_k = \sum_{j=1}^{N} h_j e^{\frac{2\pi i k t_j}{T}}$$

where $t_j$ is the transformed time of the $j^{th}$ spike, $h_j$ is the $j^{th}$ interspike interval (in transformed time), and *T* is the duration of the stimulus cycle. The dashed lines in both panels of Fig. 2C represent the mean normalized sample power of 1000 Poisson resamplings of the original spike train (NLIF on the left, Poisson on the right).[1] Our focus on the first *n* harmonics of the interval map allows us to consider features that occur no more than *n* times during the stimulus cycle—once per spike, on average. When we included more than *n* components, our ability to distinguish between NLIF and Poisson spike trains was diminished, since the power spectra of the interval maps look similar at high harmonics.

The power ratio of the NLIF spike train in Fig. 2 is 12.92, and of the Poisson-resampled spike train, 0.80. In order to say whether each spike train could have been generated by a Poisson process, we calculate the power ratios of a large number of Poisson resamplings, each of which has exactly the same PSTH as the original spike train. We consider a spike train to deviate significantly from the Poisson expectation if its power ratio is larger than the power ratios of 95% of the Poisson-resampled spike trains. For the NLIF spike train of Fig. 2, the deviation was highly significant ($p<0.0001$), while for the Poisson-resampled spike train, not surprisingly, it was not.

In Fig. 2D, we show the power ratio of NLIF spike trains as a function of stimulus contrast, as well as the mean and 95% confidence region of the power ratios from 1000 Poisson resamplings at each contrast. It is clear that at the four highest depths of modulation, 32% and above, the NLIF spike trains are readily distinguished from spike trains generated by Poisson processes with the same PSTH, and the spike trains become less and less Poisson-like as the contrast increases.

Of course, it is not necessary to calculate the power ratios in order to distinguish between the transformed-time interval maps of Fig. 2B. The Poisson interval map plainly differs from the NLIF interval map not only in the lack of clusters—the main feature captured by the power ratio—but in many other ways as well, including the shape of the summed ISIH on the vertical axis (exponential for the Poisson spike train, peaked for the NLIF spike train). In

---

[1] While one might have expected the mean power spectrum of a Poisson interval map to be flat, it actually has a low-frequency cutoff. This is because the interspike intervals (on the vertical axis) determine the values of successive spike times (on the horizontal axis), so the interval map is weakly correlated, even for a Poisson spike train. The unnormalized power at the $k^{th}$ harmonic of a Poisson interval map depends explicitly on the firing rate, and it can be shown to have an expected value of

$$|H_k|^2 = \frac{2}{(rT)^2} \left( \frac{(2\pi k / rT)^2}{1 + (2\pi k / rT)^2} \right)$$

As *k* increases, the power grows toward an asymptotic value of $2/(rT)^2$, so the frequency dependence of the power spectrum is most prominent at low harmonics (Fig. 2C).

fact, the ISIH of the NLIF data, measured in transformed time, resembles much more closely the interval distribution of a high-order gamma process. However, we choose to focus on the power ratio because, as we shall see, it distinguishes SMRPs as a class from other modulated renewal processes. Thus, the power ratio would distinguish the NLIF spike train in Fig. 2 even from a spike train generated by a high-order gamma process with a similar PSTH and ISIH (see below, *Specificity of the power ratio*, and Fig. 5).

*Sensitivity of the power ratio*. We investigated the behavior of the NLIF model with different amounts of input noise. The stimuli were 4.2 Hz sinusoidal currents at 10 contrasts, ranging from 0% to 100%. The magnitude of the response at the driving frequency was largely insensitive to the input noise (Fig. 3A). The PSTH at high contrast, however, depended significantly on the amount of input noise. Spike times were highly precise and reproducible when the input noise was low, which is reflected in a peaked PSTH (Fig. 3B, top panel). When the input noise was high (Fig. 3B, bottom panel), the PSTH peaks disappeared, indicating that in this situation the noise dominated the deterministic input and that the spike times were no longer precise and reproducible. However, even with high input noise, the power ratio distinguished high-contrast responses from Poisson spike trains (Fig. 3C).

The contrast-dependence of the deviation of NLIF spike trains from Poisson spike trains is also seen in Fig. 4A, which shows the fraction of 25 independent NLIF spike trains that had power ratios outside the Poisson range, as a function of contrast, for the same three noise levels. When the noise was low, the spike trains were either always Poisson-like or always inconsistent with Poisson processes—hence the jump from 0 to 1 between 16% and 32% contrast for the shot size of 0.0001. As the noise increased, spike trains of intermediate contrast sometimes were consistent with Poisson processes, and sometimes not. But even at the highest noise level, the responses to 100%-contrast stimuli were almost always inconsistent with Poisson processes. In other words, the power ratio distinguished Poisson spike trains from non-Poisson spike trains even when the PSTH showed no evidence of precise spike times.

*Specificity of the power ratio*. We now show that typical SMRPs cannot be empirically distinguished from Poisson-resampled spike trains by the power ratio. In Fig. 5, we present the interval maps in real and transformed time for spike trains generated by several models. Again, the marginal distribution along the horizontal axis (plotted above the interval map) is the PSTH, and the marginal distribution along the vertical axis (plotted to the right of the interval map) is the ISIH. Recall that in transformed time, the PSTH is flat by construction. The power ratio is listed in the right column, along with the *p* value from our multiple-resamplings significance test. We consider power ratios with *p* values less than 0.05 to indicate that a spike train deviated significantly from the Poisson expectation. Fig. 5A again shows an NLIF spike train, which had a highly significant power ratio. Fig. 5B shows the same NLIF response transformed into a modified Poisson process by the exchange-resampling procedure (see Methods). Fig. 5C shows the spike train of a fourth-order gamma process, and Fig. 5D shows the spike train of a $16^{th}$-order gamma process, where the firing probabilities in each case were derived from the PSTH of the original NLIF spike train. In all three cases (Figs. 5B-D), the power ratio was well within the Poisson range ($p>>0.05$), indicating that the power ratio cannot distinguish between different SMRPs, even ones with different summed ISIHs.

These results are summarized in Fig. 4B, where we plot the fraction of spike trains inconsistent with SMRPs as a function of the stimulus contrast, for several different SMRP types. Again, we simulated 25 responses at each of 10 contrasts, and the firing probability for each condition was set to match the PSTH for an example of the NLIF model with a shot size of 0.0004. At all contrasts, SMRP spike trains were empirically indistinguishable from Poisson spike trains, and hence from one another, by the power ratio. Since gamma-distributed spike trains have a relative refractory period, in which the duration of the refractory period is deterministically related to the strength of the input, we have also shown that the presence of a refractory period in itself does not produce a power ratio distinguishably different from Poisson.

Models that contain fixed refractory periods measured in units of real time, however, are not SMRPs. Such models are reasonable on biophysical grounds, because absolute refractory periods are thought to result from fundamental properties of neurons' membranes and ion channels, independent of external stimuli. Because these refractory periods are measured in real time, as explained above, they are distorted non-uniformly by the time transformation, which is determined by the overall modulation of a neuron's response and thus varies during the response. When the firing probability is low, the transformation compresses time so that the transformed-time refractory period is very short. Conversely, when the firing probability is high, the transformation expands time and thus stretches the transformed-time refractory period. Therefore, the minimum interspike interval in transformed time varies throughout the response, which leads to a modulated transformed-time interval map. These modulations may be picked up by the power ratio, which may fall outside the Poisson range.

In Fig. 6, we again present, in both real and transformed time, interval maps derived from the same NLIF spike train. We also show data from artificial spike trains generated by modulated renewal processes with fixed absolute refractory periods. The firing probability in each case was derived from the observed PSTH of the original NLIF spike train. The gray area at the bottom of the interval maps in Figs. 6B-D represents the duration of the refractory period in real (left column) and transformed (right column) time, corresponding to interspike intervals that were disallowed. In transformed time, as expected, the duration of the refractory period varied during the course of the response in all three cases. Fig. 6B shows the spike train of a Poisson process with a 2 ms refractory period, an appropriate duration for retinal ganglion cells (Berry and Meister, 1998). The power ratio for this spike train was well within the SMRP range ($p>>0.05$) despite the presence of the physiological refractory period. Fig. 6C shows the spike train of a Poisson process with a 16 ms refractory period, which is excessively long for a real neuron. In this case, the overall firing rate fell significantly, and the PSTH barely resembled the PSTH of the original spike train (Fig. 6A) because of the limitations imposed on the maximum firing rate by the refractory period. Indeed, because the refractory period was so long, it would have been impossible to obtain a PSTH identical to that of the original data. It is not surprising that in this case the power ratio was well outside the SMRP range, but it is perhaps surprising that it was still so much lower than the power ratio of the NLIF spike train. Finally, in Fig. 6D, we present the response of a $16^{th}$-order gamma process with a 2 ms refractory period, which was shorter than the typical relative refractory period of the gamma process itself. Despite the presence of both absolute and relative refractory periods, the power ratio was within the SMRP range.

The results for the refractory period models are summarized in Fig. 4C. Again, for each model, we created 250 spike trains, 25 at each of 10 contrasts. The firing probability for each condition, derived from the PSTH of the NLIF model with a shot size of 0.0004, dictated the target modulation of the firing rate. However, the degree to which the target modulation was achieved varied with the duration of the refractory period, as discussed above. Only when the refractory period was very long—above 16 ms—did the power ratios deviate significantly from the SMRP range. However, as mentioned above, such refractory periods are unreasonably long for visual neurons of the types studied here.

*Neuronal responses.* We recorded from 12 cats and three monkeys. In the cats, we measured the responses of retinal ganglion cells (recorded as S-potentials in the LGN) and LGN relay neurons. In many cases, we were able to record the responses of the LGN neurons together with their predominant retinal inputs. In the monkeys, we measured the responses of neurons in the primary visual cortex (V1). Our stimuli were drifting sinusoidal gratings of several contrasts at fixed spatial and temporal frequencies, although for a few of the retinal ganglion and LGN cells, we held the contrast fixed at 40% and varied the spatial frequency. Contrasts were logarithmically spaced, so that in most experiments the contrast was less than 40%. In general, the neuron's entire classical receptive field was stimulated by the drifting gratings, but in a few cat neurons we stimulated the center of the receptive field in isolation while keeping the surround illumination fixed at the same mean luminance.

*Retinal ganglion cells.* Altogether, we recorded 342 spike trains from 39 retinal ganglion cells. Sample interval maps from four neurons, in both real and transformed time, are shown in Fig. 7. In Fig. 7A, we show data from an X-type, on-center retinal ganglion cell. The power ratio, listed in the right column, was well outside the SMRP range, as it was for the X-type, off-center retinal ganglion cell in Fig. 7B. Note the presence of clusters of spikes in the transformed-time interval maps in Figs. 7A-B, in particular the large final cluster. These are quite similar to the clusters of spikes seen in the transformed-time interval maps of NLIF spike trains (Figs. 2, 5, 6), which suggests that phenomena similar to the reset and leak of the NLIF model may underlie spike generation in X-type retinal ganglion cells. For the Y-type retinal ganglion cell responses shown in Figs. 7C-D, the power ratio was also outside the SMRP range, but NLIF-like clusters are not obvious. While the differences between X-cell and Y-cell interval maps may be due to differences in the underlying spike generating mechanisms, we believe that the more likely explanation is that Y-cells' responses to drifting gratings typically involve a prominent elevation of the mean firing rate, with (at high spatial frequencies) a smaller modulated component than X-cells' responses (Enroth-Cugell and Robson, 1966). The transformed-time interval map, and by extension the power ratio, is highly sensitive to interactions between the modulated component of the response and spike train dynamics. Since Y-cell responses to drifting gratings are less modulated than X-cell responses, the interaction of this modulation with the dynamics of spike generation may be less obvious in Y-cells.

*LGN relay neurons*. We recorded 322 spike trains from 36 LGN neurons. Four examples are shown in Fig. 8. The power ratios for the spike trains of Figs. 8A-C were well outside the SMRP range, and the interval maps are, accordingly, highly non-uniform. For the data in Fig. 8A, recorded from an X-type, on-center LGN neuron, the transformed-time interval map seems to have several horizontal bands at the beginning (Funke and Worgötter, 1997), a single band in the middle, and the hint of a broad NLIF-like cluster at the end. Furthermore, at a transformed time of 100 ms, the mean interval becomes abruptly longer. A similar interval map is seen for the Y-cell of Fig. 8C, which also has a non-SMRP power ratio. These transformed-time interval maps differ strikingly from the transformed-time interval maps of NLIF and X-type retinal ganglion cell spike trains.

However, the response of the X-type, off-center LGN cell of Fig. 8B conforms more closely to the NLIF expectation. In this response, evoked by a 16.9 Hz, 100%-contrast drifting sinusoidal grating, there was typically only one spike per cycle. This can be inferred from the real-time interval map, which contains a prominent band at an interval of roughly 60 ms (corresponding to one spike per cycle), as well as much fainter bands above and below it (corresponding, respectively, to two spikes per cycle and one spike every other cycle). In transformed time, the interval map consists of a single, downward-sloping NLIF-like cluster. This distinct non-uniformity, which is not a simple consequence of the fact that there was typically only one spike per cycle (Poisson-resampled spike trains with the same PSTH do not have this non-uniformity), induced a remarkably high power ratio of 19.87, well outside the SMRP range.

Of the 36 LGN neurons, 31 were recorded simultaneously with their retinal input, accounting for 276 spike trains at each recording site (the number of stimulus conditions for each neuron was not identical). Interval map and power ratio analysis showed concordant power ratios in 81% of the cases. On the other hand, 43 (16%) of the paired spike trains were inconsistent with SMRPs in the retina but not the LGN, while 10 (4%) were inconsistent with SMRPs in the LGN but not the retina. An example is shown in Figs. 7D and 8D, where the interval maps are derived from the responses of a Y-type, off-center LGN neuron and its retinal input. Although the real-time PSTHs for the two cells were similar, the power ratios and transformed-time interval maps were not. In particular, the response of the retinal ganglion cell was not consistent with an SMRP, while the response of the LGN neuron was. This was not simply due to the fact that the retinal ganglion cell response had more spikes than the LGN response: when we calculated the power ratio of a shortened retinal ganglion cell spike train with the same number of spikes as the LGN spike train, the power ratio of the retinal ganglion cell spike train was still outside the SMRP range. Whether this result signifies a fundamental change in the underlying spike generating mechanism from the retina to the LGN, or whether it reflects a difference similar to the difference between X-cells and Y-cells in the retina, cannot be determined from our study. In general, our results confirm that the dynamics of LGN responses do not simply reflect their retinal inputs (Mukherjee and Kaplan, 1995).

We also recorded 113 spike trains from 19 macaque V1 neurons, of which four examples are shown in Fig. 9. In all four spike trains—two from simple cells and two from complex cells—the power ratio was significantly outside the SMRP range. The transformed-time interval maps of both simple cell responses (Figs. 9A-B) show no clear evidence of NLIF-like clusters, but rather reveal a dense but non-uniform band of points along the bottom margin. These points are likely to correspond to bursts of spikes fired within a few milliseconds of one another. Since the intervals between burst spikes are stereotyped in real time, they become variable in transformed time. The transformed-time interval map is therefore non-uniform as well, causing the power ratio to fall outside the SMRP range. Thus, the presence of bursts, which are prominent in cortical cells and are sometimes thought to convey stimulus-related information (DeBusk et al., 1997), is indicative of an underlying spike-generating mechanism that is not an SMRP.

The complex cell in Fig. 9C did not fire spikes in clear, stereotyped bursts. For this cell, the primary non-uniformity in the transformed-time interval map occurs near 25 ms in transformed time and is sufficient to elevate the power ratio outside the SMRP range. The complex cell of Fig. 9D, by contrast, had a power ratio outside the SMRP range but no obvious explanation for the modulation. This last cell had an extremely-high firing rate (around 125 imp/s), and the interval map was constructed from over 2400 spikes. The large amount of data gave rise to an extremely reliable estimate of the local interspike interval distributions, so that even small deviations from uniformity were likely to be picked up by the power ratio. Thus, even though the power ratio for this response was only 1.91, it was significantly outside the SMRP range.

Results across all recordings are summarized in Table 2. At each recording site, the fraction of spike trains inconsistent with SMRPs decreased twofold from the retina to the cortex. The decrease from retina to LGN was significant by a $\chi^2$ test (1 d.o.f., $p<0.001$), while the decrease from LGN to V1 was not significant. However, comparing retina and LGN responses, which were recorded in cats, with cortical responses, which were recorded in monkeys, is tenuous at best.

We also calculated the fraction of cells at each recording site that fired at least one spike train inconsistent with an SMRP. Because multiple spike trains were collected from each neuron, we used Bonferroni's correction (Bland, 1995) to avoid the possibility that one of the responses was significant by chance alone. Thus, if we collected $m$ spike trains from a given cell, we required that at least one of those spike trains have a power ratio outside the SMRP range with a $p$ value of $0.05/m$. With this conservative criterion, we found that 67% of retinal ganglion cells, 47% of LGN cells, and 37% of cortical cells fired at least one non-SMRP spike train (Table 2).

At all three recording sites, the fraction of spike trains that fell significantly outside the SMRP range depended strongly on the stimulus contrast (Fig. 4D). Responses of real neurons to high-contrast stimuli were more often inconsistent with SMRPs than responses to low-contrast stimuli, just as they were for the NLIF and long-refractory-period models. Since our stimulus set was heavily weighted toward low contrasts—67% of our stimuli had contrasts of 40% or less—the Bonferroni correction likely resulted in an overly conservative calculation of the number of cells that fired non-SMRP spike trains. We therefore performed a second analysis restricted to stimuli that had contrasts greater than 40%, which typically reduced $m$, or the number of responses per cell, by a factor of three. Judging from these high-contrast responses, a somewhat higher proportion of neurons at each site—and nearly 50% in the cortex—fired non-SMRP spike trains (Table 2).

*Statistics and utility of the power ratio*. The power ratio method that we have described does not rely on the use of sinusoidal stimuli or steady-state responses—it could equally well have been applied to spike trains evoked by repeated, transiently presented stimuli. What is surprising is that a large number of spikes are not required to obtain a useful estimate of the power ratio: 200 to 300 spikes, distributed over at least 16 cycles, were often sufficient to indicate the presence of a non-SMRP response. Thus, the number of spikes required to apply this method is comparable to the number of spikes required to estimate the PSTH. We do note, however, that the power ratios of non-SMRP responses increase as more and more cycles are added. This is because these power ratios detect non-uniformities in the transformed-time interval maps that are reinforced by the spikes in the additional cycles. In this sense, the power ratio actually measures the signal-to-noise ratio of a spike train's deviation from SMRP dynamics. Although a dataset of only 200 to 300 spikes may be sufficient to indicate that a spike train is inconsistent with an SMRP, longer datasets provide greater sensitivity.

## Discussion

We have presented a powerful method for distinguishing spike trains generated by two broad classes of models. The first class, simply modulated renewal processes (SMRPs), includes modulated Poisson and gamma processes. The spike-generating mechanisms in this class of models are characterized by the simple manner in which they are affected by an external stimulus: the stimulus acts simply by changing the firing rate, or, equivalently, by modulating the running speed of an internal clock. The time transformation that we employ is the unique map that regularizes the clock and thus transforms spike trains generated by these models into renewal processes.

The second class of models is characterized by underlying spike-generating mechanisms for which the effect of an external stimulus is not equivalent to a modulation of the internal clock. Such models are not SMRPs because they contain parameters that are measured in units of real time that do not covary with the stimulus. The effects of these real-time parameters survive the time transformation. The NLIF model falls into this class because it resets after each spike is fired and because it is "leaky." These features are reflected in the clusters of points at specific locations in the transformed-time interval maps of its spike trains (Fig. 2B). Models with refractory periods that are fixed in real time (Berry and Meister, 1998) also fall into this class because our time transformation distorts the refractory period, changing the duration of the refractory period nonuniformly through the response. This distortion induces a modulation in the transformed-time interval maps. If the modulation is large enough, which occurs when the refractory period is long, it is reflected in the power ratio. Our simulations in Fig. 6 show that the refractory period needs to be quite long—unphysiologically long—in order to evoke a power ratio outside the SMRP range.

The power ratio statistic was designed to be sensitive to a particular kind of structure in the transformed-time interval maps—namely, deviations from the uniformity expected of an SMRP. Many features of the interval map that could distinguish among different SMRPs do not affect the power ratio at all. For example, modulated gamma processes of different orders have different ISIHs, which are characterized by different means and variances. Their transformed-time interval maps are all uniform throughout the stimulus cycle, but the shape of the local interval distributions depends on the order of the gamma process.

Certain spike trains generated by non-SMRP models have power ratios in the SMRP range. Examples include NLIF spike trains evoked by low-contrast stimuli as well as spike trains generated by modulated renewal processes with fixed refractory periods in the physiological range. Furthermore, the power ratio is insensitive to serial correlations provided that the serial correlations are stimulus-independent. It is likely that indices other than the power ratio could distinguish these spike trains from SMRP spike trains, which suggests that our test is conservative.

Despite the lack of sensitivity of the power ratio, a surprisingly large fraction of spike trains at all three recording sites were inconsistent with SMRPs. The fraction of neurons at each recording site that had underlying spike generating mechanisms that were not SMRPs was also surprisingly large, and it was even larger when only high-contrast responses were considered. These results suggest that SMRPs are, in general, poor models for neurons in all three brain areas, especially if we believe that the underlying spike-generating mechanisms are relatively constant from one neuron to the next.

For cat X-type retinal ganglion cells in particular, several findings suggest that the NLIF model provides a useful reduced description of the spike-generating mechanism. First, such a model "phase locks" in response to sinusoidal input that is sufficiently strongly modulated (Knight, 1972), just as real retinal ganglion cells do. Second, when the stimulus modulation depth is sufficiently high, even in the presence of significant noise, evidence of the phase locking can still be seen in the PSTH; this is also true for real retinal ganglion cells (Reich et al., 1997). Third, as the contrast increases, both model and real responses undergo a gradual transition from firing spike trains that are consistent with SMRPs to firing spike trains that are not. Finally, the transformed-time interval maps of both real and model responses contain prominent resetting clusters.

Thus, the NLIF model provides a single explanation for many of the salient features of retinal ganglion cell spike trains, including details of their temporal behavior. If multiple responses of a single NLIF neuron are considered to be interchangeable with individual responses of multiple, parallel NLIF neurons (Knight, 1972), then our results may provide an explanation for the response synchronization that has been seen across multiple retinal ganglion cells (Meister et al., 1995). It should be noted, however, that our NLIF model provides a highly simplified description of retinal ganglion cell spike generation, and we made no explicit attempt to fit it to any particular retinal ganglion cell. Furthermore, our model does not contain many of the features of full-fledged models, such as realistic ion channels, and it does not account for the serial correlations between consecutive interspike intervals that are a well-described feature of unmodulated retinal ganglion cell spike trains (FitzHugh, 1958; Levine, 1991; Troy and Robson, 1992).

Because we were able to record from three successive stages of visual processing, our results also give us some insight into changes in the temporal properties of spike trains as information is transmitted through the visual system. This may be of some value in addressing the yet-unknown mechanisms of cortical information processing. The data presented in this paper suggest that retinal ganglion, LGN relay, and V1 neurons contain intrinsic temporal structure that is a consequence of their distinctive spike-generating dynamics.

# Table 1

| Model | SMRP? | Power ratio in SMRP range? |
|---|---|---|
| Noisy, leaky integrate-and-fire (NLIF) | No | No (except for low contrasts) |
| Poisson-resampled spike train | Yes | Yes |
| Modified Poisson spike train | Yes | Yes |
| Modulated Poisson process | Yes | Yes |
| Modulated gamma process | Yes | Yes |
| Non-leaky integrate-and-fire | Yes | Yes |
| SMRP with fixed absolute refractory period | No | Yes (except for excessively long refractory periods) |

**Table 1.** *Models considered in this paper.* For each model, we ask two questions: (1) Is the model a simply modulated renewal process (SMRP)? and (2) Does the power ratio fall in the range expected for an SMRP?

# Table 2

| RECORDING SITE | SPIKE TRAINS | | CELLS | | | |
| --- | --- | --- | --- | --- | --- | --- |
| | | | All Stimuli | | High-Contrast Stimuli | |
| | *Total* | *Non-SMRP* | *Total* | *Non-SMRP* | *Total* | *Non-SMRP* |
| *Retina* | 342 | 130 (38%) | 39 | 26 (67%) | 31 | 21 (68%) |
| *LGN* | 322 | 76 (24%) | 36 | 17 (47%) | 28 | 14 (50%) |
| *V1* | 92 | 21 (19%) | 19 | 7 (37%) | 19 | 9 (47%) |

**Table 2.** *Summary of results from all three recording sites (cat retinal ganglion, cat LGN, and monkey V1).* For each recording site, we tabulate the fraction of spike trains that had power ratios outside the SMRP range. We also tabulate the fraction of cells that fired such non-SMRP spike trains in response to drifting-grating stimuli of a wide range of contrasts, most of which were below 40%, and also in response to stimuli of high contrast (greater than 40%) alone.

## **Figure Legends**

**Figure 1.** *Time transformation.* Results from 128 cycles of the response of a noisy, leaky integrate-and-fire (NLIF) model (shot size 0.0004) to a 4.2 Hz sinusoidal input current at 100% contrast. **A**: response in real time (left panel) and transformed time (right panel). In the middle of each panel is a raster plot that shows the spike times in each cycle, which are collected in 1 ms bins to form the peri-stimulus time histograms (PSTHs) shown above the raster plots. When the spike times on the left are scaled by the integral of their PSTH, we obtain the demodulated, "transformed-time" version of the spike train (right panel), for which the PSTH is flat. Evenly spaced tick marks in real time (bottom of left panel), are separated non-uniformly by the time transformation (bottom of right panel)—that is, the distance between adjacent ticks is expanded when the response is strong, in the middle of the cycle, and contracted when the response is weak, early and late in the cycle. The apparent discrepancy between the number of tick marks in the two panels is due to the fact that many of the transformed-time tick marks fall on top of one another; **B**: another view of the time transformation for this spike train. The thick solid line, equivalent to the integral of the real-time PSTH, shows the value of transformed time to which each value of real time is mapped. The thin line along the diagonal represents the null transformation, in which transformed and real times are identical. When the slope of the thick solid line is greater than 1, the transformation expands time, and when the slope is less than 1, the transformation contracts time.

**Figure 2.** *The power ratio.* We use the same spike train as in Fig. 1. In panels A-C, the left column shows data taken from the NLIF model, and the right column shows the same data resampled so that the underlying statistics are those of a modulated Poisson process (see Methods). **A**: interval maps in original time; **B**: interval maps in transformed time. The large arrow in the left panel of Fig. 2B represents the resetting that occurs during the silent period of the response to each cycle, and the small arrows represent small-scale resets that occur within the response to each cycle (see Results). The marginal distributions are also shown—the PSTH along the horizontal axis and the ISIH along the vertical axis; **C**: sample power at each harmonic normalized by the total non-DC sample power. The solid line represents the normalized sample power of the test data, while the dashed line represents the mean normalized sample power at each harmonic for 1000 Poisson-resamplings of the data. The first *n* harmonics, where *n* is the smallest integer larger than the mean number of spikes in each response cycle, are signified by circles. Filled circles

bracket the cell's firing rate. Because the preponderance of the sample power for the NLIF spike train occurs in the first $n$ non-DC harmonics, we calculate the ratio of the mean sample power in the first $n$ harmonics to the mean sample power in all non-DC harmonics; **D**: power ratio of the NLIF spike train as a function of stimulus contrast, and mean and 95% confidence band for the power ratio of 100 Poisson-resampled spike trains at each contrast.

**Figure 3.** *Behavior of the noisy, leaky integrate-and-fire model as a function of input noise.* **A**: overall response, as a function of contrast, measured as the magnitude of the Fourier component at the driving frequency (4.2 Hz), for three different noise levels; **B**: PSTHs for three different noise levels at 100% contrast. Peaks in the PSTH, very sharp when the input noise is low, disappear when the input noise becomes large; **C**: power ratios for the same three different noise levels as in panel A, calculated as a function of stimulus contrast. The power ratio can distinguish the responses at all three noise levels from Poisson spike trains with the same PSTH, as long as the contrast is sufficiently high. The solid horizontal line represents the mean Poisson expectation for the power ratio across all datasets. Note that the power ratio is smaller when the input noise is larger, suggesting that NLIF spike trains with large input noise are more Poisson-like.

**Figure 4.** *Summary of analysis of model and real spike trains.* We measured the fraction of spike trains with non-SMRP power ratios for a variety of model and real neurons, as a function of stimulus contrast. **A**: NLIF models at different noise levels (shot sizes); **B**: gamma processes of different orders; **C**: Poisson processes with refractory periods of different durations; **D**: retinal ganglion, LGN, and V1 neurons. Note that stimuli of higher contrast tend to evoke non-SMRP spike trains in real neurons and in NLIF and long-refractory-period models. However, when the refractory period is in the physiologic range (on the order of a few milliseconds), higher contrasts do not cause the spike trains to deviate reliably from the SMRP expectation, suggesting that the addition of a fixed absolute refractory period to a Poisson process does not adequately account for the firing patterns of real neurons.

**Figure 5.** *Interval maps, histograms, and power ratios for four different spike-generating models.* The interval maps are presented in both real and transformed time. Modified Poisson responses are generated by the exchange-resampling procedure, while gamma responses are generated from estimates of the PSTH (see Methods). The power ratio for each response and its significance level are shown in the right column. **A**: NLIF model (noise level 0.0004);

**B**: modified Poisson process; **C**: fourth-order gamma process; **D**: 16[th]-order gamma process. Note that all of the responses derived from SMRPs have power ratios in the Poisson range and are therefore indistinguishable by this index.

**Figure 6.** *Interval maps, histograms, and power ratios for models with fixed absolute refractory periods.* Again, interval maps in both real and transformed time are shown. **A**: NLIF model (noise level 0.0004); **B**: Poisson process with 2 ms refractory period; **C**: Poisson process with 16 ms refractory period; **D**: 16[th]-order gamma process with 2 ms refractory period. The gray area at the bottom of each interval map for the refractory-period models covers the range of disallowed interspike intervals. The minimum interspike interval is fixed in real time but variable in transformed time. This means that refractory-period models are not SMRPs, even though an unphysiologically long refractory period (here, 16 ms) is required to push the power ratio outside the SMRP range.

**Figure 7.** *Interval maps, histograms, and power ratios for four different cat retinal ganglion cells.* The stimuli were all drifting sinusoidal gratings of optimal spatial frequency, and the recordings were made from S-potentials in the LGN. **A**: X-type, on-center cell, 100% contrast, 4.2 Hz; **B**: X-type, off-center cell, 100% contrast, 4.2 Hz; **C**: Y-type, on-center cell, 100% contrast, 16.9 Hz; **D**: Y-type, off-center cell, 40% contrast, 4.2 Hz. The spike train of panel D was the dominant retinal input to the LGN cell of Fig. 8D.

**Figure 8.** *Interval maps, histograms, and power ratios for four different cat LGN neurons.* The stimuli were again drifting sinusoidal gratings of optimal spatial frequency. **A**: X-type, on-center cell, 100% contrast, 4.2 Hz; **B**: X-type, off-center cell, 100% contrast, 16.9 Hz; **C**: Y-type, on-center cell, 75% contrast, 10.6 Hz; **D**: Y-type, off-center cell, 40% contrast, 4.2 Hz. The response in panel D was driven primarily by the spike train of Fig. 7D.

**Figure 9.** *Interval maps, histograms, and power ratios for four different macaque monkey V1 cells.* The stimuli were again drifting sinusoidal gratings of optimal spatial frequency. **A**: simple cell, 100% contrast, 8.4 Hz; **B**: simple cell, 100% contrast, 8.4 Hz; **C**: complex cell, 100% contrast, 4.2 Hz; **D**: complex cell, 100% contrast, 16.9 Hz.

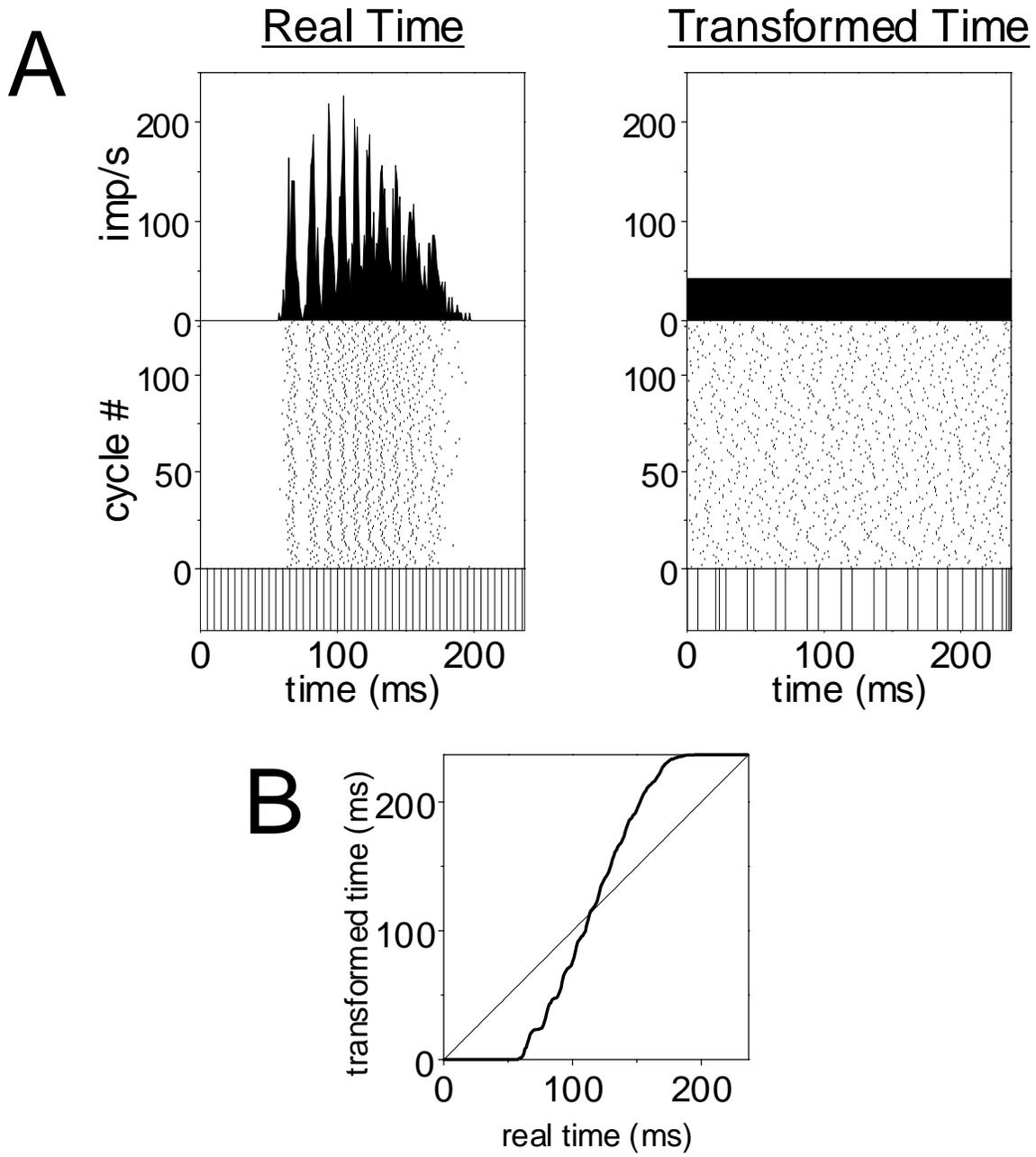

**Figure 1**

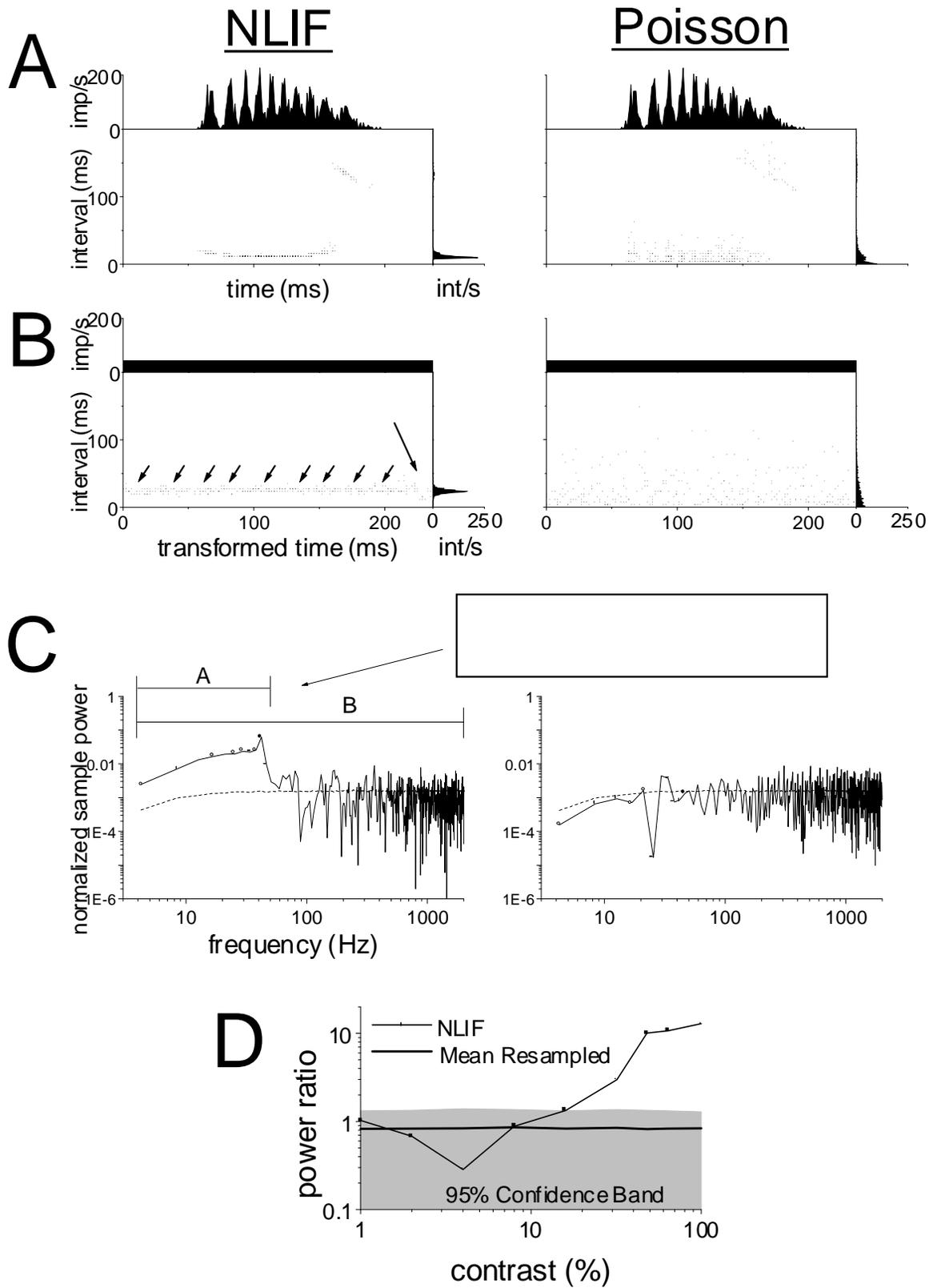

**Figure 2**

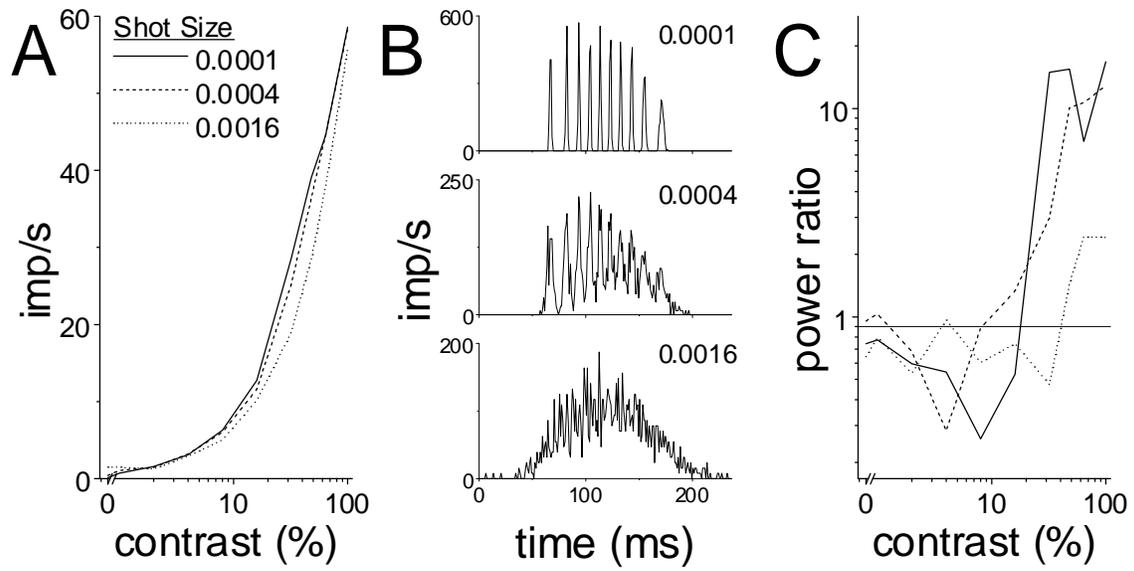

**Figure 3**

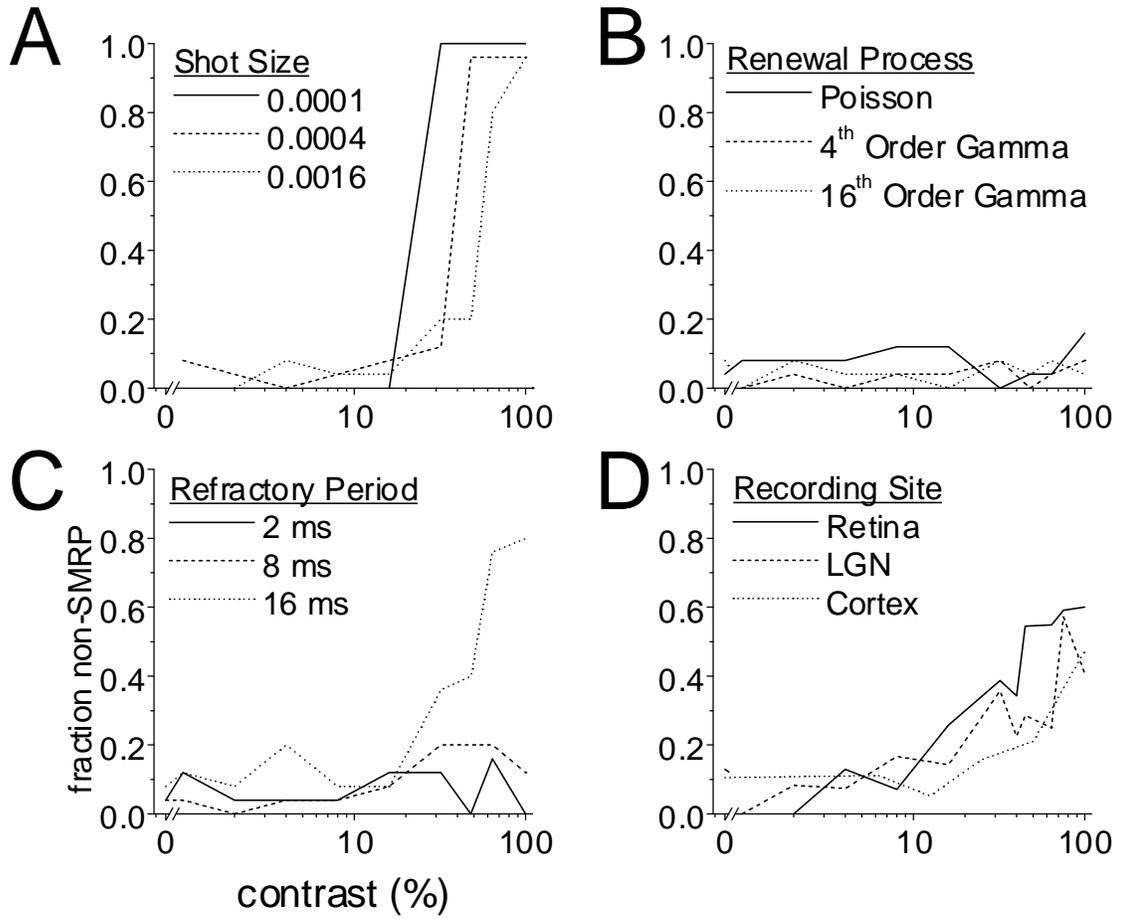

**Figure 4**

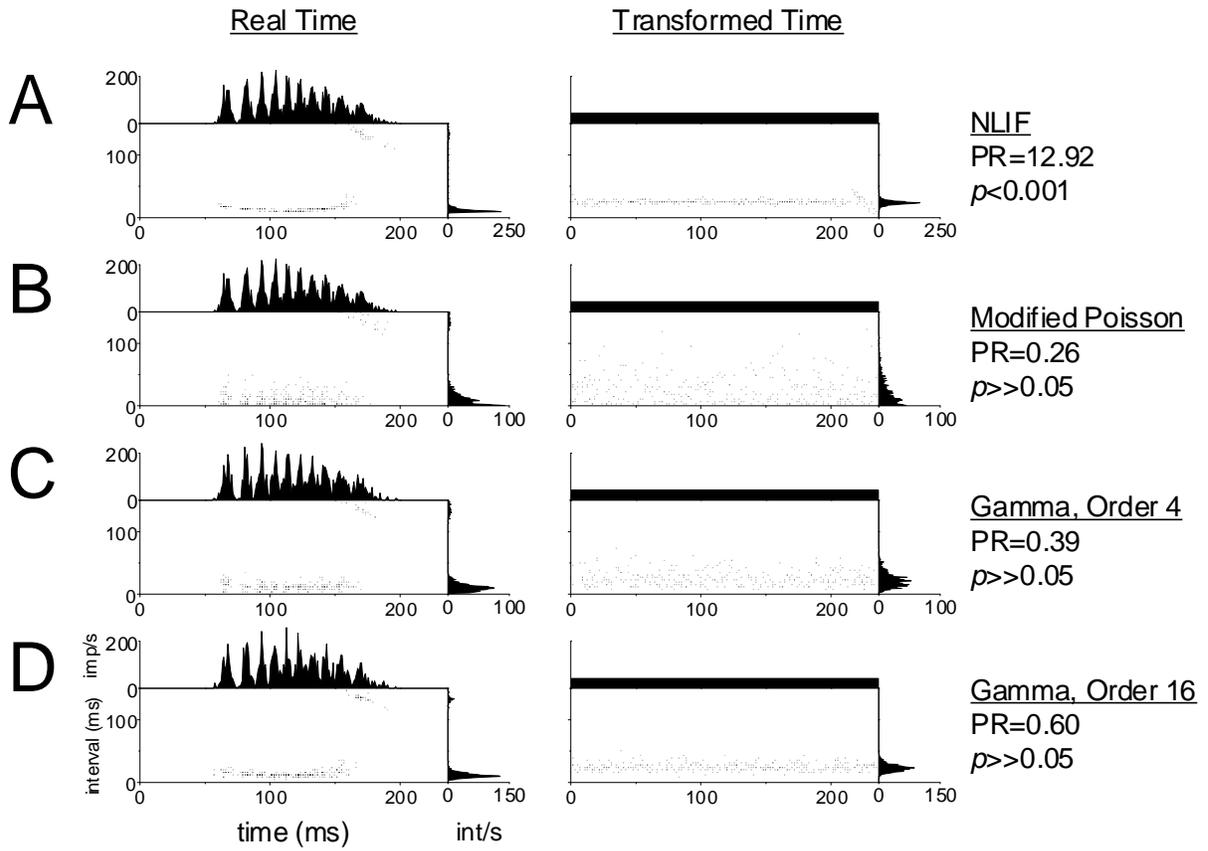

**Figure 5**

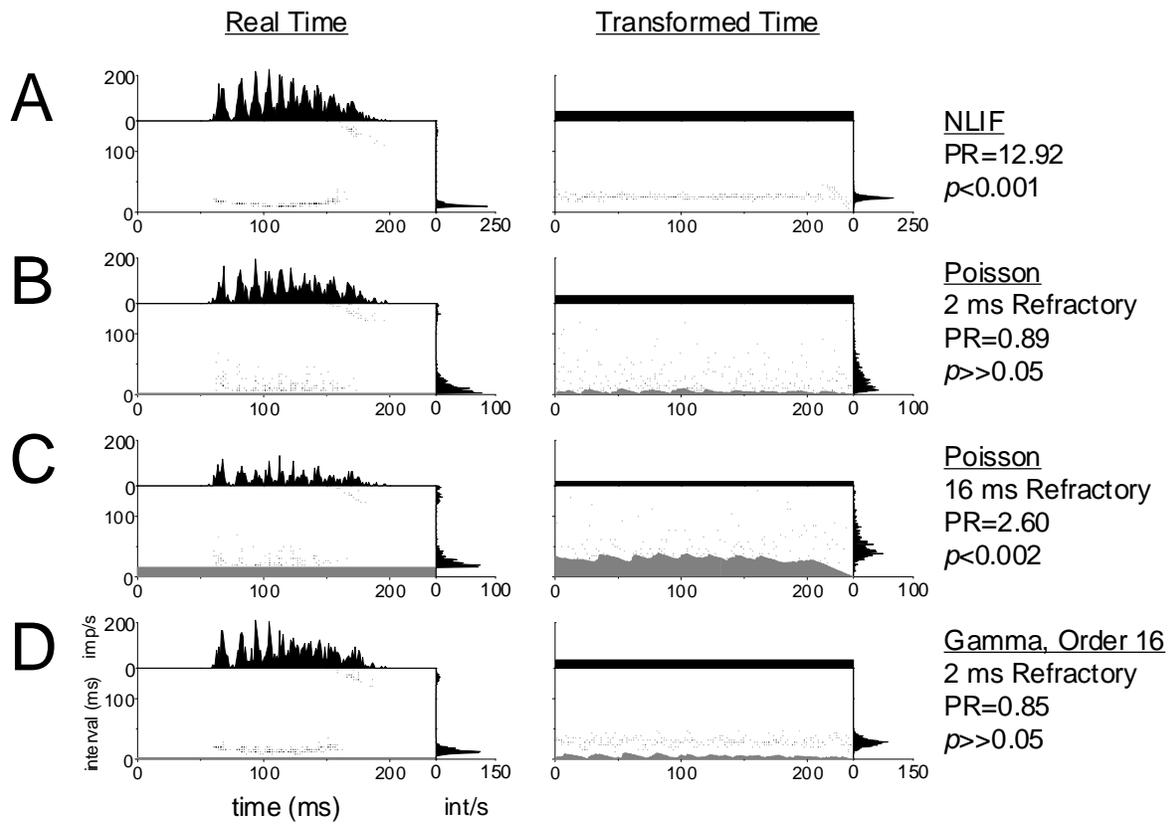

**Figure 6**

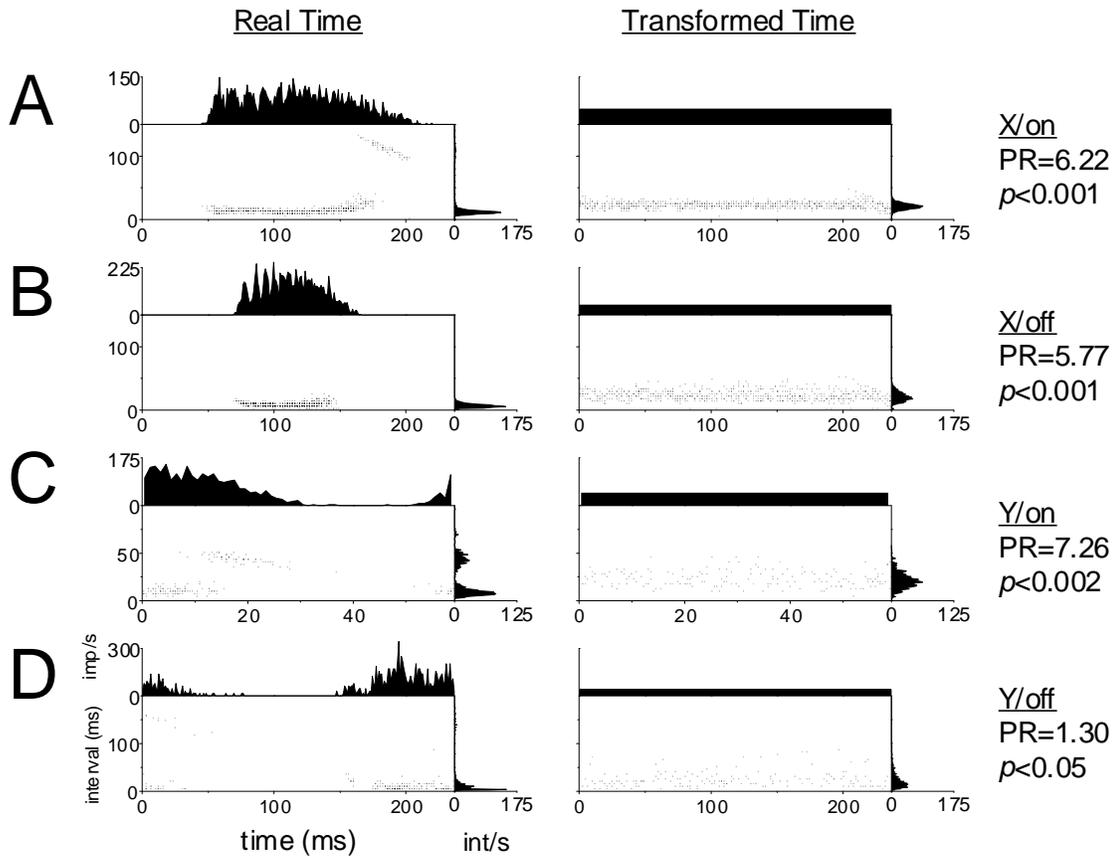

**Figure 7**

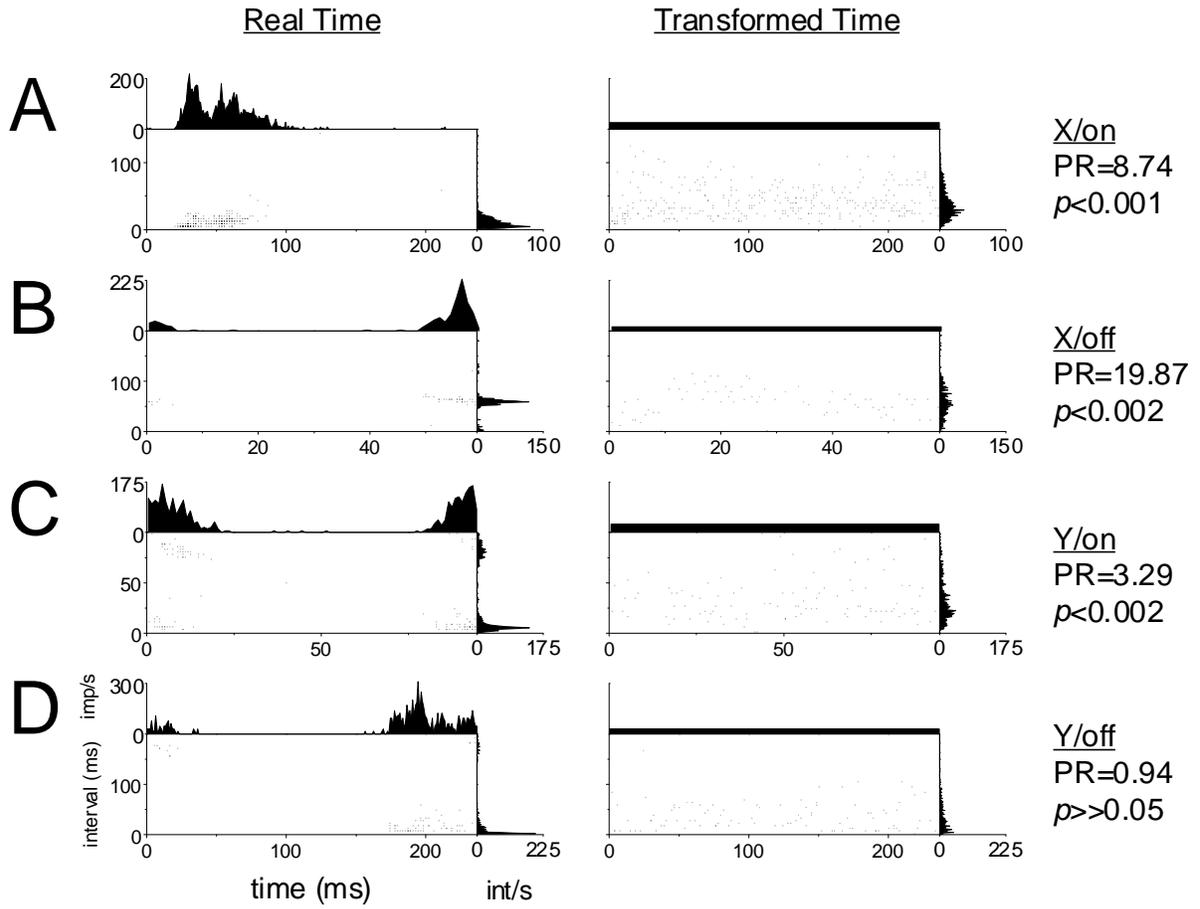

**Figure 8**

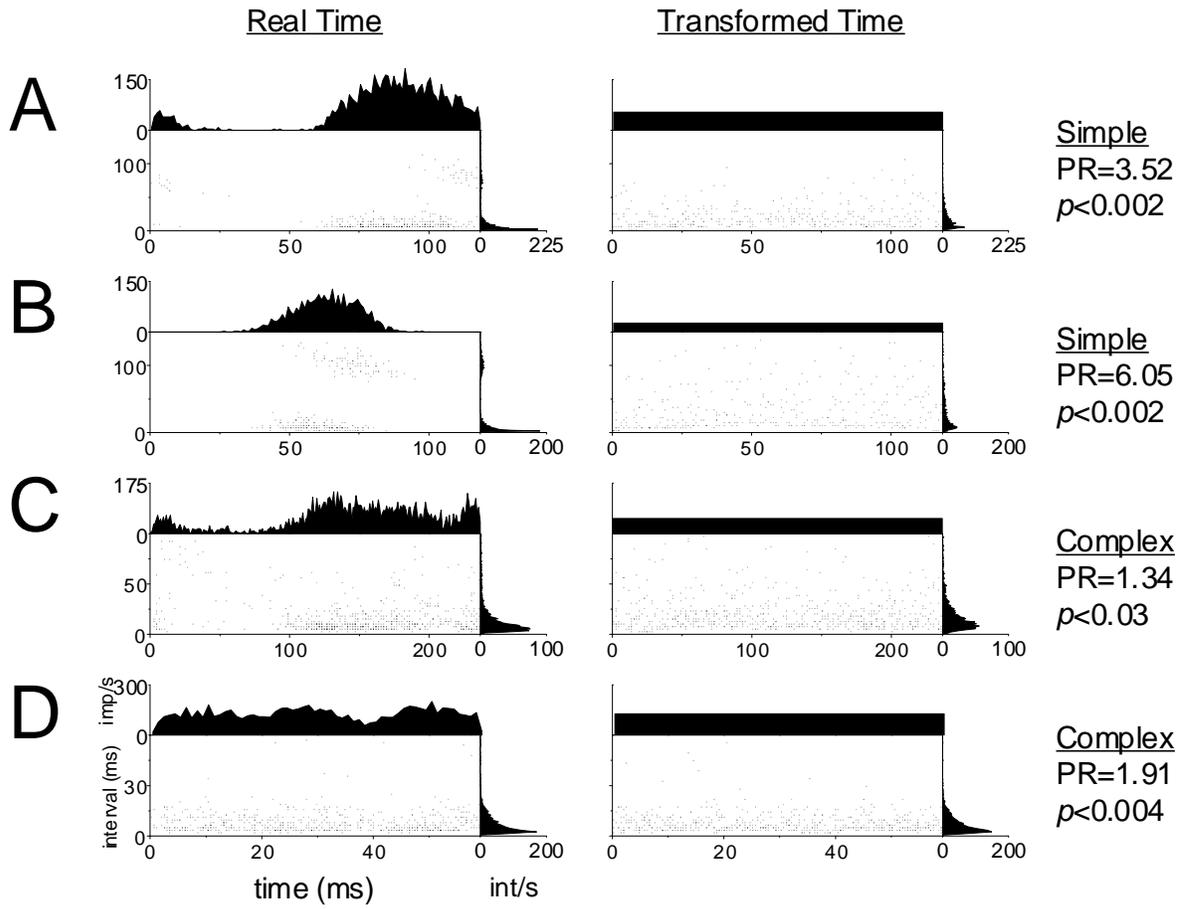

**Figure 9**